\documentclass[conference]{IEEEtran}
\usepackage{float}
\IEEEoverridecommandlockouts
\usepackage{cite}
\usepackage[numbers]{natbib}
\usepackage[colorlinks=true, linkcolor=blue, citecolor=blue, urlcolor=blue]{hyperref}

\usepackage{amsmath,amssymb,amsfonts}
\usepackage{algpseudocode} 
\usepackage{algorithm}
\usepackage{graphicx}
\usepackage{subcaption}
\usepackage{textcomp}
\usepackage{xcolor}
\usepackage{titlesec}
\titleformat{\section}
  {\normalfont\Large\bfseries\filright}
  {}                                  
  {0pt}                               
  {}                                  

\def\BibTeX{{\rm B\kern-.05em{\sc i\kern-.025em b}\kern-.08em
    T\kern-.1667em\lower.7ex\hbox{E}\kern-.125emX}}
\begin{document}

\title{\huge{Analyzing and Optimizing the Distribution of Blood Lead Level Testing for Children in New York City:\\ A Data-Driven Approach}}

\author{\IEEEauthorblockN{Mohamed Afane and Juntao Chen}
\IEEEauthorblockA{Department of Computer and Information Sciences,
Fordham University,
New York, United States \\
Email: \{mafane,jchen504\}@fordham.edu}
}

\maketitle

\noindent \textbf{Abstract}   This study investigates blood lead level (BLL) rates and testing among children under six years of age across the 42 neighborhoods in New York City from 2005 to 2021. Despite a citywide general decline in BLL rates, disparities at the neighborhood level persist and are not addressed in the official reports, highlighting the need for this comprehensive analysis.
In this paper, we analyze the current BLL testing distribution and cluster the neighborhoods using a k-medoids clustering algorithm. We propose an optimized approach that improves resource allocation efficiency by accounting for case incidences and neighborhood risk profiles using a grid search algorithm. Our findings demonstrate statistically significant improvements in case detection and enhanced fairness by focusing on under-served and high-risk groups.
Additionally, we propose actionable recommendations to raise awareness among parents, including outreach at local daycare centers and kindergartens, among other venues.
\bigskip

\noindent \textbf{Keywords} Blood Lead Levels in Children, Clustering, Grid Search

\section*{Introduction}

Lead is a toxic heavy metal with no biological role in the human body, and exposure to it can cause severe health issues, including learning disabilities and behavioral problems \cite{NYCHealth2023}, \cite{Hauptman2017}. It often accumulates from deteriorating lead-based paint, contaminated dust, soil, water, and no level of lead in the blood is considered safe \cite{Naranjo2020}, \cite{Jones2009}, \cite{Vorvolakos2016}, \cite{Leighton2003}.

Urban areas, older housing, and high-risk groups such as low-income and non-Hispanic black children are disproportionately impacted\cite{jacobs2002}. And despite legislative and educational efforts to mitigate lead exposure, the challenge remains persistent among these vulnerable communities \cite{Hauptman2017}, \cite{Jones2009}.

In children, lead exposure is a critical public health issue, leading to serious and sometimes irreversible neuro-developmental damage \cite{Hauptman2017}, \cite{Aizer2018}. Vulnerable to its effects, children can accumulate lead, disrupting cellular functions and impairing brain development. Even low-level exposure can have profound, albeit sometimes subtle, neurological  consequences \cite{Naranjo2020}.

The research emphasizes that there is no safe level of lead exposure, and even minimal amounts can lead to developmental deficits. These effects can include symptoms like poor attention, memory loss, kidney damage, headaches, muscle tremors, and hallucinations in Acute cases. Other effects include fatigue, muscle weakness, insomnia, joint pain, gastrointestinal issues, and reduced performance in standardized tests \cite{assi2016}, \cite{flora2006}, \cite{Papanikolaou2005}. An extensive study that was conducted in 2008 comparing BLL and children's performance on IQ tests at different stages of their development concluded that the children's intellectual ability was impaired by BLLs that are even lower than 10 micrograms per deciliter (mcg/dL), a unit measuring the concentration of lead in the blood \cite{Jusko2003}. A similar study found that an increase of 1 mcg/dL in lifetime BLL was associated with a 0.87 decrease in IQ \cite{Canfield2003}.
Another 2017 study concluded that each lead-exposed child has an annual cost of \$5600 in medical and special educational expenses, while the US loses \$50.9 billion in economic productivity due to cognitive impairments associated with Lead exposure \cite{Hauptman2017}.

In 2004, New York City passed a law that requires the New York City Department of Health and Mental Hygiene (DOHMH) to report the progress regarding the city's efforts towards reducing BLL among children in NYC \cite{NYCHealth2023}. Section I of this report focuses on children under 6 years of age, and that is the focus of this paper. Section IV of the report is also important for our purpose as it highlights the current testing efforts, stating that most children in NYC were actually tested for lead at least once before they turned 3. The exact number as of the 2023 report is 77 percent. This is very relevant as it shows that about 23 percent of children were never tested for lead, which is over 100,000 children \cite{NYCHealth2023}. 
Identifying the most vulnerable areas can give us a good idea of how to efficiently distribute tests in the future. Even though other factors like health insurance and family income can have important effects on testing, identifying the areas that need more testing is very relevant for future testing initiatives conducted by the city or other no-profit organizations.
COVID-19 had a very important effect on testing in NYC, as the total number of tests decreased yearly after 2019, which is another reason to efficiently allocate the limited resources that we have to detect more cases and help the most vulnerable areas.

While the existing body of research and reports mentioned above examined this problem from the health perspective mainly, previous studies have not fully addressed two critical factors that are central to our research:

\begin{itemize}
    \item \textbf{Recency}: No recent study has examined lead exposure data in New York City, making it crucial to understand current trends and developments. Our study fills this gap by providing an updated analysis.
    
    \item \textbf{Testing Focus}: Prior research has often overlooked the importance of testing distribution. Simply identifying areas with high lead levels is insufficient; it’s equally important to assess how thoroughly those areas are being tested. Our study uniquely combines lead rate analysis with testing data to provide a more comprehensive understanding.
\end{itemize}
By addressing these gaps, our study offers valuable insights that are essential for developing more effective lead testing and intervention strategies in New York City.

\section{Methods}
The data used in this study is publicly available and can be found on the New York Environment and Health Data Portal \cite{NYCLeadData}, As of now the data is available up to 2021 and it has information on the number of tests conducted every year in each neighborhood, with the number of cases and rates for 3 levels; 15+ mcg/dL, 10+ mcg/dL, and 5+ mcg/dL. This last feature will be our focus for the rest of the study.
As the environmental intervention threshold is lowered over time, the new data might include 3.5+ mcg/dL or even lower levels \cite{NYCHealth2023}. 

In this section, we analyze the BLL rates across neighborhoods over time. Yearly scaling is crucial to understand trends relative to annual averages. While a Min-Max scaler, commonly used to enhance machine learning performance and mitigate outlier effects normalizes data by adjusting each value between a predefined minimum and maximum \cite{Ahsan2021}, \cite{Singh2020}. It may not be ideal for our dataset where outliers play a critical role in the analysis. Instead, mean normalization is employed, as it is more suited to our needs. This method scales data points around a value of 1, facilitating an intuitive understanding of each point's deviation from the average. The mean normalization formula is given by:

\begin{equation}
    x_{\text{normalized}} = \frac{x_i}{\text{mean}(x)}
\end{equation}

where \(x_i\) is a data point in the series, and \(\text{mean}(x)\) is the average of the series \(x\). Normalized values are expressed as 1, less than 1, or greater than 1, indicating whether data points are equal to, below, or above the average, respectively.
\bigskip

\begin{figure}[H]
\centerline{\includegraphics[scale=0.41]{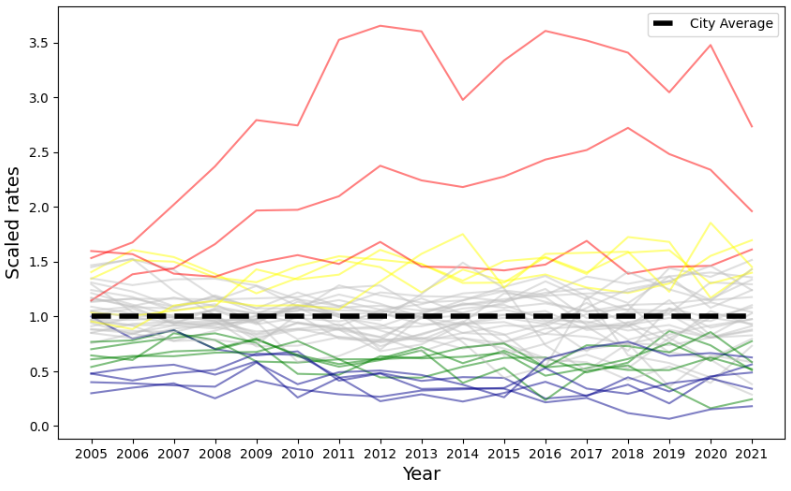}}
\caption{Rates for each neighborhood in NYC scaled.}
\label{Figure 1}
\end{figure}

Fig. \ref{Figure 1} visualizes the change in neighborhood-level BLL rates, illustrating how certain areas have consistently exhibited higher rates than the citywide average since data collection began. This granular view highlights the variances across different neighborhoods and underscores the critical need for localized public health interventions to address these disparities effectively
 
\subsection{\textbf{Identifying the Level of Vulnerability for Each Neighborhood}}

Our study employs the k-medoids algorithm, which was chosen for its interpretability, as it uses actual neighborhoods as cluster centers. K-medoids is also effective for clustering time-series data, ensuring that our analysis captures both current and historical trends in BLL rates \cite{Radovanovic2020}. The steps of the algorithm are:

\begin{enumerate}
    \item Select $k$ initial medoids based on domain knowledge and data analysis.
    \item Assign each data point to the nearest medoid based on similarity.
    \item Optimize the medoid of each cluster to minimize intra-cluster distances.
    \item Iterate over steps 2 and 3 until the medoids stabilize.
\end{enumerate}

For our specific application, medoid identification was guided by the following criteria based on BLL:

\begin{itemize}
    \item \textbf{High}: GeoIDs marking areas with consistently high BLL deviations, indicating significant exposure risks.
    \item \textbf{Low}: GeoIDs from neighborhoods with notably low BLLs, reflecting minimal lead exposure risks.
    \item \textbf{Average}: GeoIDs that mirror the citywide average BLL trend, representing typical urban exposure.
    \item \textbf{Rising}: GeoIDs from areas experiencing an upward trend in BLLs, suggesting increasing exposure risks.
    \item \textbf{Declining}: GeoIDs for neighborhoods showing a decreasing trend in BLLs, denoting improving conditions.
\end{itemize}

The decision to use five clusters, rather than the typical three (High, Average, and Low), stems from the need to capture key distinctions in rate trends across different areas. For example, rising clusters exhibited relatively low rates prior to 2010 but saw significant increases afterward, potentially due to recent events or environmental factors. This contrasts with consistently high-rate clusters, where the rates may be driven by long-standing issues. Similarly, declining clusters, which previously had high rates, have experienced decreases over time, distinguishing them from areas with persistently low rates. These five clusters allow for a more nuanced understanding of BLL exposure levels and trends across New York City.

\begin{figure}[H]
\centerline{\includegraphics[scale=0.38]{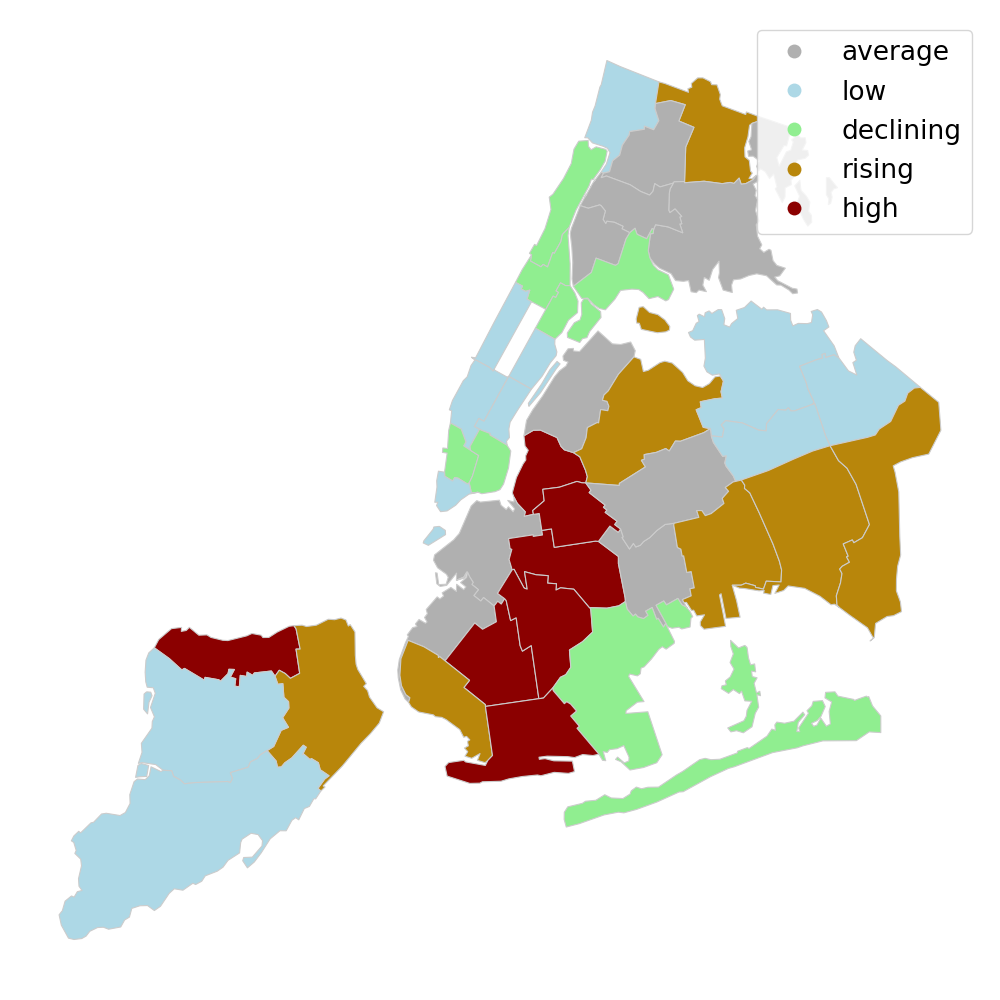}}
\caption{The final clustering.}
\label{Figure 2}
\end{figure}
The map in Fig. \ref{Figure 2} summarizes the clustering and clearly shows a geographical pattern across the city. We can observe that 85\% of the areas that maintained high rates relative to the city average are in Brooklyn, which might be due to the age of buildings, as buildings built prior to 1960 and did not have regular maintenance present a higher risk of lead exposure \cite{jacobs2002}, \cite{NYCHealth2023}. Another cause might be Tap water quality in these areas \cite{pocock1983}, \cite{watt1996}. A further study is needed to investigate the underlying reasons for the higher rates in this specific region.

A more detailed view of the clustering results, focusing on how rates change across time for the 5 identified clusters (excluding the average one), is provided in the supplementary materials online (see Supplementary Figure 1). Neighborhoods in the red cluster, representing the highest rates, and the dark yellow cluster, indicating rising rates, are particularly prominent in Brooklyn and Queens. These clusters appear darker in contexts where color distinction is less visible.

In contrast, neighborhoods in the blue and green clusters, associated with consistently lower or recently declining rates, are represented with lighter shades. These areas, while lower in risk, still require attention to ensure effective resource allocation. The distinction between "low" and "declining" clusters is particularly significant; while "low" areas have consistently shown minimal BLLs, "declining" clusters were once characterized by much higher levels, especially before 2010. This historical context underscores the need for continued monitoring in "declining" areas, as they may still harbor residual risk factors that could affect future BLLs. By understanding these distinctions, we can better tailor our optimization strategies to ensure that interventions are not only responsive to current risk levels but also informed by the trajectory of exposure in these neighborhoods.

\subsection*{\textbf{Analyzing The Current Testing Distribution}}

\begin{figure}[htbp]
\centerline{\includegraphics[scale=0.62]{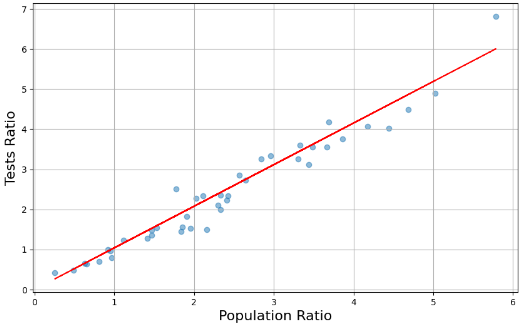}}
\caption{The testing ratio as a function of the child population.}
\label{Figure 3}
\end{figure}

When we analyze the number of tests conducted per neighborhood, the most interesting thing to notice is that distribution is mostly static and does not factor in the changing rates in each area. 
After analyzing different features, the child population under 5 years of age, seems to have the highest correlation with the testing conducted for children under 6 years.
This shows the importance of having a more dynamic testing policy that considers the needs of each neighborhood as they change. Neighborhoods might have similar child populations but different risk profiles and, therefore, different testing needs. with very few exceptions. The ratios of tests conducted in each area are similar to the starting ratios when the metrics were first reported in 2005.
Fig. \ref{Figure 3} illustrates the strong correlation between the population and the testing ratios with a slope 1.04.
This shows how static the testing policy is, as it doesn't take into account the changing rates and needs of individual neighborhoods.
 
A primary issue with having test numbers align with child population is the potential oversight of varying risk profiles. Neighborhoods with similar populations might show equal numbers of tests, despite having significantly different needs. This can lead to areas with larger populations but lower risk levels using more testing resources inefficiently.

To illustrate the inefficiency of this approach, we compare the rates, tests conducted, and cases detected between Greenpoint (UHF code=201) and South Beach (UHF code=504).

\begin{figure}[htbp]
\centerline{\includegraphics[scale=0.42]{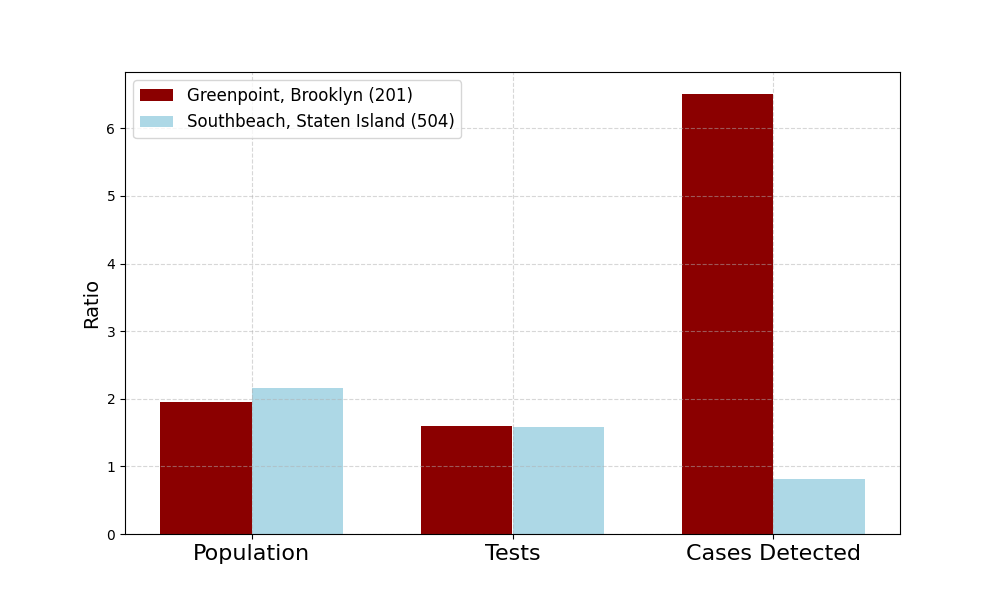}}
\caption{Metrics of Greenpoint vs South Beach.}
\label{Figure 4}
\end{figure}

Fig. \ref{Figure 4} shows that both neighborhoods have about 2 percent of the child population in New York City, giving them both a similar number of tests over the years, even though Greenpoint has been consistently averaging 8 times more cases than South Beach. 
In 2021, 3760 tests were conducted in Greenpoint, resulting in 97 discovered cases, while 3720 tests were conducted in South Beach, resulting in just 12 detected cases.

A similar pattern occurs when we compare Kingsbridge (101) and Port Richmond (501) among other cases. the analysis of these cases reveals that neighborhoods such as Greenpoint and Port Richmond rank among the most under-tested in the city, as evidenced by their elevated rates. While increasing testing frequency involves a multitude of factors, pinpointing regions with the most critical needs serves as an essential initial step.

\subsection*{\textbf{Testing Neighborhood With Low Rates}}
The second concern with the current distribution of testing is efficiency, particularly when a large number of tests are conducted in neighborhoods with low rates. A more effective strategy would prioritize regions with higher BLL rates to optimize public health responses. For instance, Lower Manhattan (310) saw only 2 positive cases from 1,200 tests in 2021, despite conducting 5,500 tests since 2018 to find only 7 cases. This starkly illustrates the inefficiency in using resources given the low incidence rate.

Redirecting tests to higher-risk areas, such as neighborhoods with older housing or near industrial zones, would enhance intervention effectiveness. This recalibration would ensure a more equitable distribution, align testing efforts with actual needs, and improve the overall public health strategy against lead exposure.

\subsection*{\textbf{Optimization Techniques}}

In our study, we employed Grid Search to refine the distribution strategy for lead testing. This method tests out all possible combinations of the given parameters, making it highly effective but computationally expensive with larger datasets \cite{Liashchynskyi2019}, \cite{Alibrahim2021}. However, given the relatively small dataset in our study, the computational cost is manageable, allowing us to comprehensively address the optimization problem with all relevant parameters while maintaining simplicity and effectivness of the apporach.

The first step in our approach involves a clear articulation of our objective, which we will translate into mathematical and programming terms. Our primary goal is to ensure an efficient and equitable distribution of testing resources. We will measure efficiency by comparing the increase in new case detection achieved by our revised strategy against the previous distribution while maintaining the same total number of tests citywide.

To ensure fairness, we will set specific limits to ensure that even areas with low or declining rates still get tested enough, and use the risk profiles and child populations of different areas to ensure that the number of tests allocated does not exceed them. Given the city's objective to test all children, approaching these population numbers is acceptable, provided the figures are not exceeded. Our strategy for optimizing resource distribution aims not just to maximize case detection mathematically but also to find a balanced distribution that adheres to both efficiency and fairness constraints. This comprehensive approach is crucial for the successful implementation of our testing strategy. These guidelines are set to ensure that resources are directed where they are most needed and that vigilance is maintained across all regions.

\subsection*{\textbf{Objective Function}}

In our approach, we define two key equations. The first one computes the new testing ratios (tests/population) for each neighborhood, which we will call v2 testing ratios, whereas the current version is called v1 testing ratio (often close to 1):
\begin{equation}
    \text{v2 tests ratio} = \sum_{i=1}^{n} \left( x_i \cdot p_1 + y_i \cdot p_2 \right),
    \label{eq:v2_tests_ratio}
\end{equation}
where \( x_i \) represents the current testing ratios  and \( y_i \) the `cases ratio` for each neighborhood \( i \) which is the percentage of cases detected in the area out of all cases detected in the city, over the past 3 years, since these are more indicative of future rates.  
The parameters \( p_1 \) and \( p_2 \) are weights applied to these ratios. The function normalizes the weighted sum of the testing and cases ratios so that the total across all neighborhoods sums to 100\%. This normalized ratio is then used to determine the number of tests for each neighborhood based on the total available tests \( T \).

The second equation calculates the difference in the number of cases due to the new testing distribution:

\begin{equation}
    \Delta C = T \times \sum_{i=1}^{n} \left( R_i \times  (\rho_{2i} - \rho_{1i}) \right),
    \label{eq:case_difference}
\end{equation}

where:
\begin{itemize}
    \item \( \Delta C \) represents the `Cases Difference'.
    \item \( T \) is the `total tests' conducted.
    \item \( n \) is the number of neighborhoods.
    \item \( R_i \) represents the `rates' for neighborhood \( i \).
    \item \( \rho_{2i} \) and \( \rho_{1i} \) represent the `v2 cases ratio' and `v1 cases ratio' for neighborhood \( i \), respectively.
\end{itemize}

Our objective is to maximize this `Cases Difference', which signifies the change in projected cases due to the adjustments in testing distribution. This optimization aims to identify the combination of \( p_1 \) and \( p_2 \) that yields the most number of cases while following the previously defined criteria.

This function is then optimized using the grid search approach over a wide range of values of \( p_1 \) and \( p_2 \). Grid Search, as defined previously, will systematically evaluate varying testing ratios \( p_1 \) and \( p_2 \), providing a thorough assessment of potential distributions. This method's strength lies in its effectiveness and exhaustive nature:

\begin{algorithm}
\caption{Optimize Testing Distribution Using Grid Search}
\begin{algorithmic}[1]
\Require Rates and tests for each GeoID, constraints on case differences, and population ratios
\Ensure Maximizing positive case differences
\State Define the function based on Equation~\eqref{eq:v2_tests_ratio} to compute the weights \( p_1 \) and \( p_2 \))
\For{each value in the range of \( p_1 \)}
    \For{each value in the range of \( p_2 \)}
        \State Calculate new \( v2 \) testing ratios
        \State Compute new \( v2 \) tests and \( v2 \) cases
        \State Compute the difference in cases using Equation~\eqref{eq:case_difference}
        \If{all constraints are met}
            \State Save the \( p_1 \), \( p_2 \) combination, and cases detected
        \EndIf
    \EndFor
\EndFor
\State Save parameters with the highest cumulative difference
\State Recalculate distribution using the best parameters
\State \textbf{return} Final optimized distribution
\end{algorithmic}
\end{algorithm}

\section*{Results}

The evaluation of the new testing distribution is done using an estimated 260,000 tests calculated by a simple regression model based on the total number of tests conducted in the past years. we then calculate the number of cases detected using the current distribution and compare it the number of cases detected using the optimized distribution.
The current model identified 2,860 cases under the existing testing distribution, whereas the optimized model projected an increase to 3,270 cases, reflecting a 14.3\% improvement in detection. To assess the sensitivity of the optimization results, we tested three different ranges of the critical parameters (p1 and p2), including broader ranges such as -10 to 10 and narrower ones like -1 to 1. Despite these variations, the results remained consistent, further supporting the robustness of the approach.

This enhancement suggests a potential annual increase of over 410 cases, predominantly in highly vulnerable areas. Statistical analysis, utilizing a Z-test for proportion comparison, confirmed the significance of this increase (\(p < 0.05\)), underscoring the strategic impact of our proposed redistribution. 

\begin{figure}[htbp]
\centerline{\includegraphics[scale=0.52]{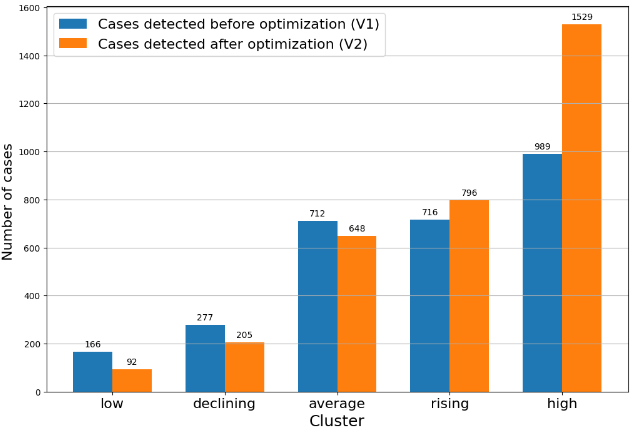}}
\caption{The total number of cases by cluster before and after the optimization.}
\label{Figure 5}
\end{figure}

Fig. \ref{Figure 5} shows the change in the number of cases detected between the two models; while our model might potentially detect fewer cases in the low-risk areas, the trade-off is well justifiable considering the number of new cases discovered in high-risk neighborhoods.
This analysis highlights the critical role of data-driven approaches in enhancing the effectiveness of public health interventions, particularly in optimizing resource allocation and improving outcomes in vulnerable communities.

\begin{figure}[htbp]
\centerline{\includegraphics[scale=0.4]{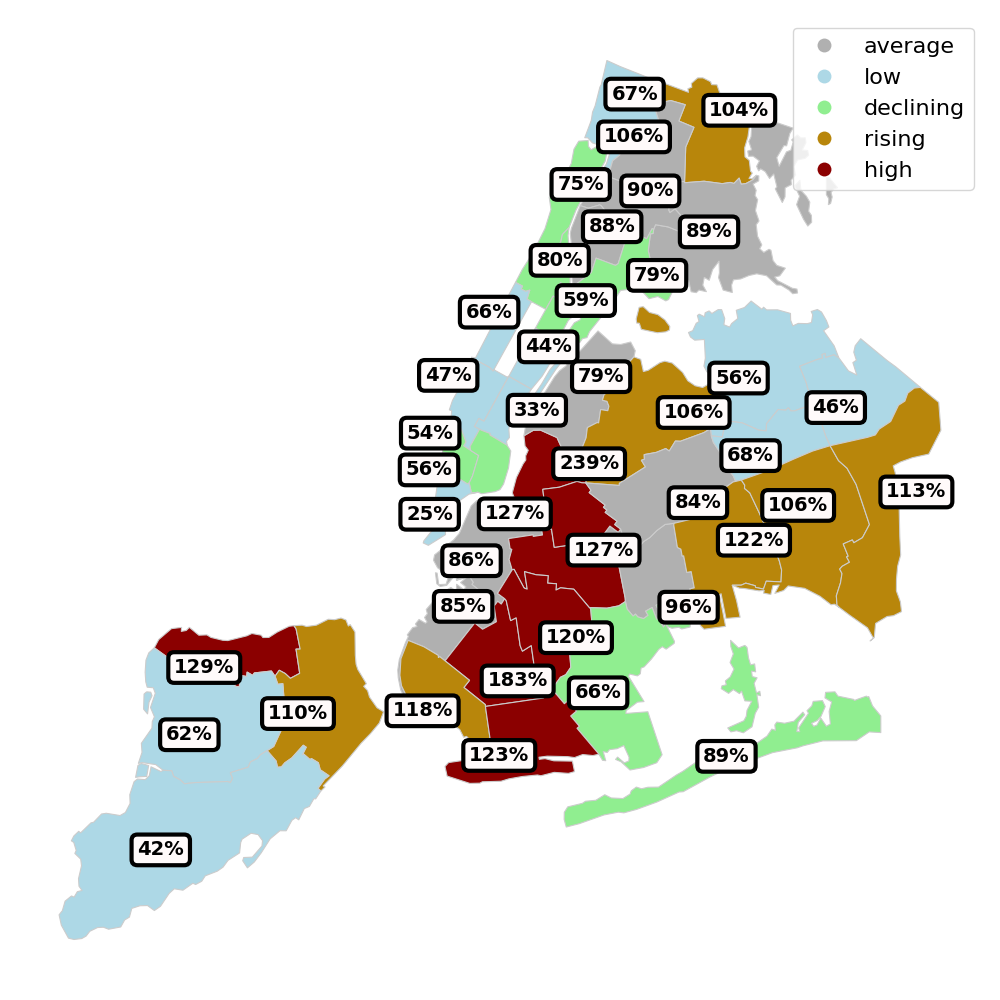}}
\caption{The testing over population before the optimization.}
\label{Figure 6}
\end{figure}

Fig. \ref{Figure 6} visualizes the impact of resource reallocation on our new model by comparing current testing distributions to previous ones. It shows the percentage of original allocations that are now required. For instance, Lower Manhattan now needs only 25\% of its former tests, enabling ongoing monitoring with greater resource efficiency. The figure highlights significant reductions in test allocations in areas with historically low or declining BLL cases, which is consistent with our strategy to prioritize higher-risk neighborhoods. 

\section*{Discussion}
We propose several recommendations that could be implemented by policymakers, healthcare providers, or other organizations committed to enhancing public health measures, particularly in monitoring and addressing childhood lead exposure in urban environments. 

\noindent\textbf{Detailed Data Tracking}

The data is available at the UHF 42 level only. To refine BLL testing strategies and interventions, we recommend the adoption of more granular data tracking methods. Specifically, BLL data should be monitored and reported at least at the zip code level or even more detailed geographic markers if possible. 
This approach would allow for the identification of micro-level trends and risk factors, enabling more targeted and effective public health responses. 

\noindent\textbf{Optimized Distribution as Standard Practice}

The proposed testing distribution shows the importance of a data-informed, equitable BLL testing strategy,
The optimized testing distribution model developed in this study could serve as a baseline for minimum testing standards in each area. By implementing this model, health departments can ensure that all neighborhoods, particularly those at higher risk, receive an adequate number of tests. This model not only maximizes resource efficiency by focusing on high-need areas but also establishes a framework for ongoing evaluation and adjustment of testing policies.

\noindent\textbf{Awareness Campaigns}

To increase BLL testing rates and public awareness, particularly in high-risk neighborhoods, we recommend the launch of targeted awareness campaigns. Public health messages should be disseminated through various channels, including local daycare centers and kindergarten schools, which represent vital platforms for educating parents and guardians about the risks of lead exposure and the importance of testing. Additionally, train lines such as B, D, F, Q, and R, which traverse many high-risk areas, could also be targeted. These campaigns should emphasize the importance of early detection.

\noindent\textbf{Water Quality and BLL Monitoring for Pregnant Women}

Given the elevated BLL rates in Brooklyn and Queens, it is crucial for local authorities to undertake comprehensive water quality evaluations. Regular testing is essential to manage any heightened lead levels effectively. Additionally, intensified monitoring of BLL of pregnant women in these areas is necessary. This approach aims to mitigate lead-related health issues and ensure safer environments for children.

\noindent\textbf{Limitations}

This study's primary limitation is the lack of more detailed BLL data within specific neighborhoods. certain zip codes might be disproportionately affected by elevated BLL rates, yet the absence of such detailed data precludes a more nuanced analysis.  
Another potential improvement can be made during the clustering and risk evaluation process. We chose the 5 clusters based on a detailed analysis of the data, but as the rates change over time, another framework might be needed to highlight the importance of ongoing research and analysis of the rates and testing for this critical issue.

\section*{Conclusion}

This study analyzed blood lead levels data for children under six years of age in New York City at the neighborhood level, focusing mainly on the current testing distribution, and the risk profiles of each neighborhood.  We introduced an optimized testing distribution using a Grid Search optimization algorithm. this distribution considers current testing ratios, current rates, and neighborhood risk profiles determined through the k-medoids clustering algorithm, aiming to enhance testing precision and fairness.
The proposed distribution resulted in a statistically significant increase in the number of cases detected, and a more equitable policy by focusing on areas with the most needs. Other recommendations were made that are cost-effective and could be implemented by public health officials, healthcare institutions, and other organizations interested in addressing this important issue.
\bigskip

\noindent \textbf{Data Availability}

The data used in this study is publicly available and can be found on the New York Environment and Health Data Portal \cite{NYCLeadData}. The codes used for optimization and data analysis are available upon request.

\bibliographystyle{unsrtnat}
\bibliography{references}   

\end{document}